\documentstyle[epsf,eqsecnum,aps,prb,twocolumn]{revtex}

\begin{document}
\draft
%\preprint{HEP/123-qed}

%%%%%%%%%%%%%%%%%%%%%%%%%%%%%%%%%%%%%%%%%%%%%%%%%%%%%
\twocolumn[\hsize\textwidth\columnwidth\hsize\csname
@twocolumnfalse\endcsname
%%%%%%%%%%%%%%%%%%%%%%%%%%%%%%%%%%%%%%%%%%%%%%%%%%%%%

\title{
Magnetic anisotropy of the spin ice compound Dy$_{2}$Ti$_{2}$O$_{7}$
}

\author{
H. Fukazawa\cite{email} 
}
\address{
Department of Physics, Kyoto University, Kyoto 606-8502, Japan 
}
\author{
R. G. Melko 
}
\address{
Department of Physics, University of Waterloo, Waterloo, 
Ontario, N2L 3G1, Canada 
}
\author{
R. Higashinaka 
}
\address{
Department of Physics, Kyoto University, Kyoto 606-8502, Japan 
}
\author{
Y. Maeno 
}
\address{
Department of Physics, Kyoto University, Kyoto 606-8502, Japan \\
International Innovation Center, Kyoto University, Kyoto 606-8501, Japan \\
and CREST, Japan Science and Technology Corporation, Kawaguchi, 
Saitama 332-0012, Japan \\
}
\author{
M. J. P. Gingras 
}
\address{
Department of Physics, University of Waterloo, Waterloo, 
Ontario,N2L 3G1, Canada \\
and Canadian Institute for Advanced Research, 180 Dundas Street West,
Toronto, Ontario, M5G 1Z8, Canada \\
}
\date{\today}
\maketitle

\begin{abstract}
We report magnetization and ac susceptibility of single crystals of 
the spin ice compound Dy$_{2}$Ti$_{2}$O$_{7}$. 
Saturated moments at 1.8 K along the charasteristic axes 
[100{\rm ]} and [110{\rm ]} agree with the expected values 
for an effective ferromagnetic nearest-neighbor Ising pyrochlore 
with local $\left<{111}\right>$ anisotropy, where each magnetic moment is 
constrained to obey the `ice-rule'. 
At high enough magnetic fields along the [111{\rm ]} axis, 
the saturated moment exhibits a beaking of the ice-rule; 
it agrees with the value expected for a three-in one-out 
spin configuration. 
Assuming the realistic magnetic interaction between Dy ions 
given by the dipolar spin ice model, we completely reproduce 
the results at 2 K by Monte Carlo calculations. 
However, down to at least 60 mK, 
we have not found any experimental evidence of the long-range 
magnetic ordering predicted by this model to occur at around 180 mK.  
Instead, we confirm the spin freezing of the system below 0.5 K.
\end{abstract}


%%%%%%%%%%%%%%
]\narrowtext
%%%%%%%%%%%%%%

\section{INTRODUCTION}

The observed absence of magnetic ordering 
in the Ising pyrochlore magnet Ho$_{2}$Ti$_{2}$O$_{7}$ had a strong impact 
on the present study of frustrated systems \cite{Har1}. 
This compound was known to have a ferromagnetic (FM) Curie-Weiss temperature 
$\theta_{\rm CW} \sim 1.9 \, {\rm K}$, and magnetic Ho$^{3+}$ ions which 
possess a strong Ising anisotropy along the local $\left<{111}\right>$ axis. 
Previous to this discovery, it was expected that absence of ordering 
would take place in the pyrochlore lattice 
only with antiferromagnetic Heisenberg spins \cite{Rei1,Can1}. 

Harris {\it et al.} showed the possibility of a macroscopically 
degenerate ground state in Ho$_{2}$Ti$_{2}$O$_{7}$ by 
explaining the multiplicity of the possible spin configurations. 
By analogy with proton ordering in ice I$_{\rm h}$ \cite{Har1,Pau1}, 
the spin configuration with two spins pointing inward and 
two spins pointing outward for each tetraheron 
corresponds to the `ice-rule' for these compounds. 
Accordingly, Ho$_{2}$Ti$_{2}$O$_{7}$ and related compounds 
such as Dy$_{2}$Ti$_{2}$O$_{7}$, which also possesses FM Ising spins 
on a pyrochlore lattice, were named `spin ices'. 

The first naive theoretical model of the spin ice materials placed
Ising spins on a pyrochlore lattice, constrained to point along the local 
$\left<{111}\right>$ direction,  with only nearest-neighbor FM exchange 
interactions \cite{Har1,Har2}. 
However, it was soon deduced that the nearest-neighbor exchange interaction 
in these compounds is actually \textit{antiferromagnetic} \cite{Sid1} and 
that the effective nearest-neighbor interaction becomes ferromagnetic 
due to the long-range dipolar interaction arising from 
the large moments of the Ho$^{3+}$ and Dy$^{3+}$ ions \cite{Her1}. 
The physical model which correctly incorporated all of these 
interactions is called the {\it dipolar spin ice} model. 
Physical quantities (neutron scattering cross sections and 
specific heat curves) calculated from Monte Carlo simulations 
of this model are in excellent agreement with 
the corresponding experimental data \cite{Her1,Ram1,Bra1}.

Most experimental studies on the spin ices to date have been performed 
on polycrystalline samples \cite{Har1,Sid1,Ram1,Bra1,Jan1,Mat1,Bra2}. 
Among these various experiments, 
the most dramatic and direct evidence for realization of the spin ice state 
is the residual entropy equal to Pauling's entropy: 
$\frac{1}{2}{\it R}\,{\rm ln} \frac{3}{2} = 1.68$ 
${\rm J/K}^{2}$mol \cite{Pau1}. 
Ramirez $et$ $al.$ obtained an entropy very close to this value 
($\sim 1.9(2)$ ${\rm J/K}^{2}$mol-Dy) by integrating the 
experimental electronic specific heat $\it {C_{e}}/{\it T}$ 
of Dy$_{2}$Ti$_{2}$O$_{7}$ above 0.25 K \cite{Ram1}. 
This residual entropy can be removed by applying an external magnetic field, 
since the magnetic field can lift the macroscopic degeneracy by stabilizing 
a specific spin configuration with lower magnetic energy (Fig. 1), 
as proven by neutron diffraction measurement on single crystals of 
Ho$_{2}$Ti$_{2}$O$_{7}$ and Dy$_{2}$Ti$_{2}$O$_{7}$ \cite{Har1,Fenn1}. 

\begin{figure}
\begin{center}
    \epsfxsize=8cm
	\epsfbox{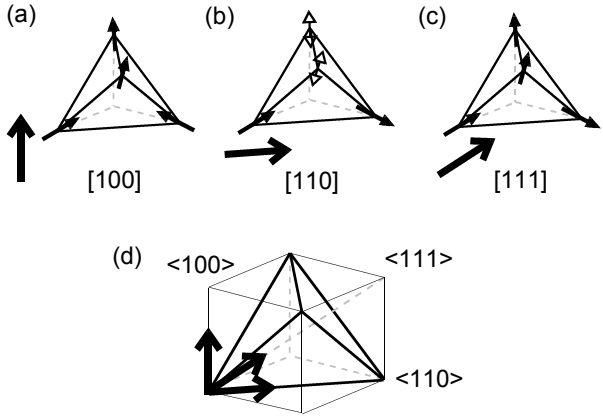}
\end{center}
\caption{
Spin configurations with the magnetic field along (a) the [100], (b) [110]
and (c) [111] axes.
For the [111] direction the figure depicts 
the state in which the ice-rule breaks. 
(d) Three dimensional axes in order to give three dimensional perspective 
of the direction in which the field is applied. 
}
\end{figure} 

Moreover, the electronic specific heat, ${\it C_{e}}({\it T})/{\it T}$, 
of Dy$_{2}$Ti$_{2}$O$_{7}$ polycrystals exhibit three peaks at 0.34 K, 0.47 K 
and 1.12 K independent of the magnitude of the magnetic field \cite{Ram1}. 
Due to the freedom of the decoupled spins of Fig. 1(b), it is speculated 
by Ramirez $et$ $al$. that these peaks are attributable to 
crystal grains with their [110] axis aligned parallel to the field. 

The above experimental facts suggest that 
the spin ice compounds have significant magnetic anisotopy, and therefore 
one must control the field direction carefully 
when studying their physical properties. 
In particular, the competition between the applied magnetic field and the 
spin-spin interactions that drive the ice-rule obeying degenerate ground state 
is significanty affected by the crystal orientation in the field . 
However, to date there has been no experimental report 
detailing the behavior of this anisotropy 
using single crystals of Dy$_{2}$Ti$_{2}$O$_{7}$. 
In this paper, we report such measurements and compare our results 
with the results predicted by the dipolar spin ice model. 
We show experimental magnetization results at 1.8 K 
along three characteristic axes of this compound, 
namely the [100], [110] and [111] axes, and 
the corresponding numerical results calculated by Monte Carlo simulations. 
As we will discuss, agreement between the two is excellent. 
In contrast, a similar set of measurements carried out by Cornelius and 
Gardner on Ho$_{2}$Ti$_{2}$O$_{7}$ failed to find the expected 
magnetic anisotropy associated with this material \cite{Cor1}. 

In addition, it has recently been theoretically predicted that 
for the dipolar spin ice model and, consequently, for Dy$_{2}$Ti$_{2}$O$_{7}$ 
and Ho$_{2}$Ti$_{2}$O$_{7}$, a first-order phase transition 
should occur at around 180 mK ($\equiv {\it T}_{\rm c}$) 
which fully releases Pauling's ice entropy \cite {Mel1}. 
Melko $et$ $al.$ predicted the transition by considering a spin loop 
move instead of single spin flips within their Monte Carlo simulation, 
as illustrated in Fig. 1 of Ref. \ref{Rog1}. 
However, few explicit experimental searches for this transition 
have been reported \cite{Fenn1,Mat2}. 
Hence, the second motivation for this paper is to investigate 
single crystals of Dy$_{2}$Ti$_{2}$O$_{7}$ 
using susceptibility measurements down to 60 mK, 
in order to search for the predicted magnetic ordering. 

It is hoped that this experimental and theoretical joint study 
on Dy$_{2}$Ti$_{2}$O$_{7}$ can help answer some current issues of debate, and 
help develop deeper understanding of the nature of the spin ice materials.

\section{Experiments and Calculations}

Single crystals of Dy$_{2}$Ti$_{2}$O$_{7}$ were grown 
by the floating zone method with an infrared furnace 
equipped with double-elliptical mirrors \cite{Nec1}. 
Before the single crystal growth we prepared a polycrystalline feed rod 
by thoroughly mixing Dy$_{2}$O$_{3}$ (99.99\%) and TiO$_{2}$ (99.99\%) 
powders with the molar ratio of 1 to 2 and sintering 
the mixture in an alumina crucible at 1000$^{\circ}$C for 100 hours in air. 
The typical growth conditions are 29.0 mm/h for the feed speed 
and 30.0 mm/h for the crystal growth speed. 
To avoid oxygen deficiency, we grew the single crystals 
in 0.3 MPa of O$_{2}$ atmosphere. 
In addition, we annealed the as-grown crystals 
at 800$^{\circ}$C for 100 hours in O$_{2}$ gas flow. 
The color of the annealed crystals is homogeneously translucent yellow, 
while that of the as-grown crystals is lighter yellow 
with some inhomogeneous color distribution. 
The sizes of samples used for magnetization and ac susceptibility measurements 
are $0.7 \times 0.7 \times 2.0$ mm$^{3}$ and 
$3.0 \times 1.0 \times 0.2$ mm$^{3}$, respectively. 
By performing powder x-ray diffraction measurements on the annealed crystals, 
we found that the scattering pattern was consistent with 
a cubic unit cell with $a =$ 10.124(5) \AA$\,$without any extra peaks. 
We determined the principal axes using Laue pictures. 

We measured magnetization between 1.8 and 40 K for fields up to 5 T 
with a SQUID magnetometer \cite{Qua1}. 
We measured ac susceptibility by a mutual-inductance method down to 60 mK 
using a dilution refrigerator \cite{Oxf1}. 

For magnetization measurements, we applied the magnetic field in the plane 
perpendicular to the longest side of the crystal, 
because all of the principal axes are contained within this plane. 
The field misalignment is estimated to be within a few degrees \cite{Qua1}. 
Thus, we estimate the demagnetization factor $n$ as 0.4 for this crystal. 
The demagnetization effect can be neglected for ac susceptibility 
measurements, since the ac field was applied parallel to the longest side, 
which is parallel to the [100] crystal axis. 

In addition, we performed Monte Carlo simulations for magnetization
at finite temperatures on the dipolar spin ice model.
The detailed calculation techniques are described in Ref. \ref{den1}.
We adopted the following Hamiltonian:
\begin{eqnarray}
H &=& -J \sum_{\langle i,j \rangle} {\textit{\textbf{S}}\,}_{i}^{z_{i}}
 \cdot{\textit{\textbf{S}}\,}_{j}^{z_{j}}
  - \sum_{i}  {\textit{\textbf{H}}\,}_{\text{eff}}
      \cdot  {\textit{\textbf{S}}\,}_{i}^{z_{i}}  \nonumber \\
  & & +D \, r_{\rm nn}^{3} \sum_{j > i} \Bigg[ \frac{{\textit{\textbf{S}}\,}_{i}^{z_{i}}\cdot{\textit{\textbf{S}}\,}_{j}^{z_{j}}}{|{\textit{\textbf{r}}}_{\it ij}|^{3}} - \frac{3({\textit{\textbf{S}}\,}_{i}^{z_{i}} \cdot 
{\textit{\textbf{r}}}_{\it ij})({\textit{\textbf{S}}\,}_{j}^{z_{j}} \cdot {\textit{\textbf{r}}}_{\it ij})}{|{\textit{\textbf{r}}}_{\it ij}|^{5}} \Bigg] . 
\end{eqnarray}
Here ${\textit{\textbf{S}}\,}_{i}^{z_{i}}$ is the Ising moment vector of 
magnitude of unity at site $i$ pointing along the local Ising axis $z_{i}$. 
The quantity ${\textit{\textbf{H}}\,}_{\text{eff}}$ is the
effective internal magnetic field which directly couples to each spin. 
For the Hamiltonian coefficients, we used $-1.24$ K for the nearest-neighbor 
exchange interaction $J_{\rm nn} \equiv J/3 $ and 2.35 K 
for the nearest-neighbor dipolar interaction $D_{\rm nn} \equiv 5D/3$. 
We took into account the long-range nature of the dipolar interactions 
through the Ewald summation technique. 
A standard Metropolis algorithm was used with a conventional cubic unit cell 
for the pyrochlore lattice containing 16 spins. 
Simulations of the specific heat obtained using these values are 
in very good agreement with the experimental specific heat 
reported for Dy$_{2}$Ti$_{2}$O$_{7}$. 
The simulation size was $L = 3$ (432 spins). 
When calculating the magnetization curves, a number of different simulation 
sizes were tried, and little variation resulted.

\section{Results and Discussion}

\subsection{Magnetic Anisotropy}

In Fig. 2 we summarize the magnetization curves at 1.8 K 
along the three characteristic axes. 
We took the horizontal axis as the effective magnetic field, 
$\mu_{0}{\it H}_{\rm eff} = \mu_{0}{\it H}_{\rm ext} - n{\it M}_{\rm exp}$. 
We obtained nearly the same data at 2 K (not shown). 
We note that no hysteresis is observed 
between the upward and downward field sweeps. 

Saturated moments along the [100] and [110] axes agree with 
the expected values for Ising anisotropy along the local $\left<{111}\right>$ 
direction and for effective FM coupling between the nearest-neighbor Dy ions. 
The value for the [100] axis is 
$g_{\it J}{\it J}(1/\sqrt{3}) = 5.77$ $\mu_{\rm B}{\rm /Dy}$ 
(two spins point in and two spins point out: Fig.1(a)) 
and that for the [110] axis is 
$g_{\it J}{\it J}\{ (\sqrt{2/3} \times 2)/4 \} = 4.08$ $\mu_{\rm B}{\rm /Dy}$ 
(one spin points in, one spin points out and two spins are free: Fig. 1(b)). 
Here, $g_{\it J}$ represents Land\'{e}'s $g$ factor, 
$\it J$ corresponds to the total angular moment of Dy$^{3+}$ ion, and 
$\mu_{\rm B}$ represents the Bohr magneton. 
For the free Dy$^{3+}$ ion, $g_{\it J}{\it J}$ is expected 
to be 10 $\mu_{B}$/Dy. 
Slight deviation from the expected values may be 
due to misalignment effect as we discuss below. 

Below 0.5 T, the magnetization along the [111] axis already 
reaches the value $g_{\it J}{\it J}\{ (1 + 1/3 \times 1)/4 \} = 3.33$ 
$\mu_{\rm B}{\rm /Dy}$ expected for the saturated state 
in which the ice-rule completely holds. 
In this state, three of the spins per tetrahedra are coupled to minimize 
their energy with the applied magnetic field along the [111] direction, 
but one spin chooses to oppose the field and minimize its energy with respect 
to the effective nearest-neighbor interaction, remaining in the ice-rule. 
For quite large fields (${\it H}_{\rm eff} >> J_{nn}+D_{nn}$), 
this configuration begins to cost a large amount of energy, and 
the field-opposed spin chooses to flip, minimizing its energy 
with respect to the applied magnetic field, but breaking the ice-rule. 
Accordingly, the magnetization agrees with the value 
expected for the three-in one-out spin configuration (Fig. 1(c)), 
$g_{\it J}{\it J}\{ (1 + 1/3 \times 3)/4 \} = 5.00$ $\mu_{\rm B}{\rm /Dy}$. 
We infer that the gain of Zeemann energy exceeds 
that of the effective nearest-neighbor FM interaction above 0.5 T: 
$(1/3)g_{\it J}{\it J}\mu_{\rm B}(\mu_{0}{\it H}_{\rm eff}) = 1.12\,{\rm K}$, 
in good accordance with ${\it J}_{\rm eff} \equiv 
{\it J}_{\rm nn} + {\it D}_{\rm nn} = 1.11\,{\rm K}$. 

\begin{figure}
\begin{center}
    \epsfxsize=8cm
	\epsfbox{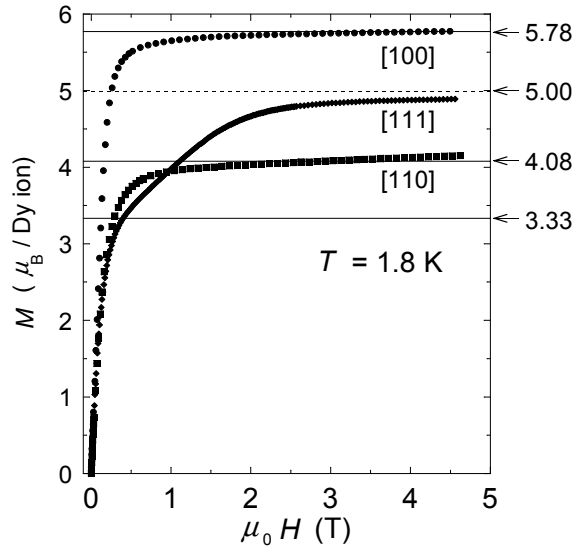}
\end{center}
\caption{
Magnetization curves at 1.8 K along the [100], [110] and [111] axes. 
Solid lines correspond to the expected saturated moments 
for the two-in two-out spin configuration for all the direction: 
dotted line corresponds to that for the three-in one-out spin configuration. 
}
\end{figure}

For fields up to 0.5 T, 
the moments along all the axes nearly reach the expected values 
for the corresponding spin-ice configurations in a magnetic field. 
This experimental fact indicates that the magnetic anisotropy 
in Dy$_{2}$Ti$_{2}$O$_{7}$ is indeed compatible with the spin-ice model. 
In addition, the breaking of the ice-rule observed along the [111] axis 
confirms that the strong Ising anisotropy 
along the local $\left<{111}\right>$ direction dominates in this compound. 

A corresponding field dependence of magnetizations is 
reported by Cornelius and Gardner 
for single crystals of Ho$_{2}$Ti$_{2}$O$_{7}$ \cite{Cor1}. 
However, their data above 2 T is totally unexpected, 
because the magnetizations along all the directions 
reach approximately 5.9 $\mu_{\rm B}{\rm /Ho}$. 
Their behavior cannot be explained solely by a crystal misalignment 
effect, since the effect of misalignment exists in our experiment 
but results in different behavior \cite{Yas1}. 

In Fig. 3 we show magnetizations at 1 and 2 K calculated 
by the Monte Carlo simulations. 
Each solid symbol represents the magnetization along the corresponding field 
direction of the experimental results. 
The open squares are for the magnetization at 2 K along the [110] axis 
with intentional 0.5 degree misalignment toward the [100] axis. 
The open diamonds are for the magnetization at 1 K along the [111] axis. 

Clearly the Monte Carlo simulations at 2 K reproduce 
our corresponding experimental magnetization curves very well, though 
the absolute values do not completely coincide with each other \cite{Not1}. 
This excellent agreement between experiments and calculations implies that 
one may treat this compound as an ideal model system 
for the study of geometrical frustration in the pyrochlore lattice.
 Moreover, it shows that the assumption of an Ising doublet 
with leading wavefunctions $\vert \psi_+\rangle$ and  $\vert \psi_-\rangle$ 
with $\vert m_J\rangle$ decompositions 
$\vert \psi_\pm\rangle \approx \vert m_J=\pm 15/2\rangle$ 
is a very good approximation, 
in good agreement with crystal-field calculations for Dy$_{2}$Ti$_{2}$O$_{7}$ 
\cite{rosenkranz-private} adapted from the crystal-field parameters 
extracted for Ho$_{2}$Ti$_{2}$O$_{7}$ \cite{rosenkranz-neutron}. 

\begin{figure}
\begin{center}
    \epsfxsize=8cm
	\epsfbox{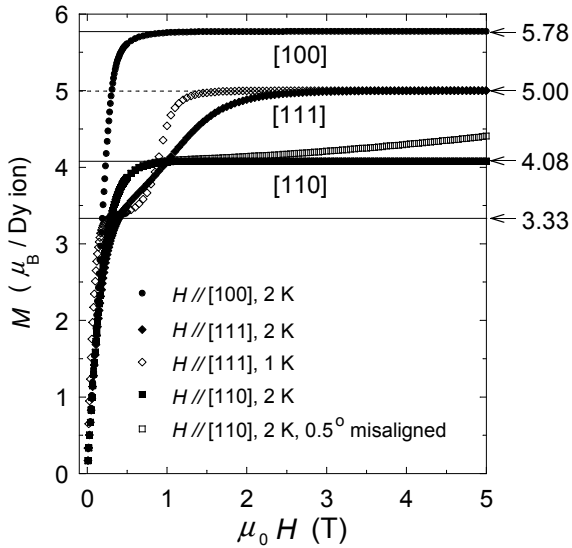}
\end{center}
\caption{
Magnetization curves calculated by Monte Carlo simulations 
of the dipolar spin ice model. 
The simulation size was $L = 3$ (432 spins), and parameters 
were chosen for Dy$_{2}$Ti$_{2}$O$_{7}$. 
}
\end{figure}

The calculated magnetization with the field along the [110] axis 
with the intentional 0.5 degree misalignment deviates substantially 
from the aligned value at higher fields. 
This is due to the acute angular sensitivity 
specific to field application along the [110] axis, 
since slight deviation of the field from the [110] axis can fix 
the spin directions of the 2 free spins normal to the [110] axis 
in every tetrahedron (Fig. 1(b)). 
The increase in magnetization is attributable to the field-coupled component 
of these spins still satisfying the ice-rule. 
Therefore, the experimental deviation from the expected value, 
4.08 $\mu_{\rm B}{\rm /Dy}$, can be explained by assuming 
a crystal misalignment within 0.5 degree from the [110] axis. 

The derivative of the experimental and calculated magnetizations at 1.8 K 
with ${\it H}_{\rm eff}$ along the [111] axis are shown in Fig. 4.
The derivative of the experimental magnetization, 
$d{\it M}_{\rm exp}/d{\it H}_{\rm eff}$, exhibits a definite kink 
at around 0.5 T and quite a broad peak at around 1 T. 
This kink signals the crossover from two-in, two-out 
spin-ice configuration to three-in, one-out configuration. 
The derivative of the calculated magnetization, 
$d{\it M}_{\rm cal}/d{\it H}_{\rm eff}$, also exhibits 
a local minimum at around 0.5 T and a broad peak at around 1 T. 
In addition, the behavior below 0.5 T nearly coincides with 
the experimental results. 
This good agreement between experiments and calculations indicates that 
the behavior of the dipolar spin ice model in the spin-interacting regime 
({\it i. e.} at low fields) behaves almost the same as the real materials 
in the same regime. 
The difference between them is most likely due to the experimental 
misalignment effect, since the difference became smaller when we measured 
another single crystal by adjusting the alignment more accurately \cite{Not2}. 

\begin{figure}
\begin{center}
    \epsfxsize=8cm
	\epsfbox{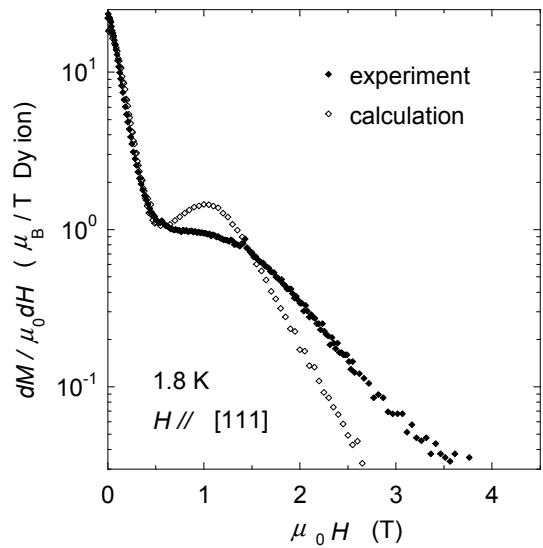}
\end{center}
\caption{ 
Derivative of the experimental and calculated magnetizations at 1.8 K 
with $\mu_{0}{\it H}_{\rm eff}$ along the [111] axis. 
}
\end{figure}

The calculated broad peak becomes sharper at 1 K and 
correspondingly a clear plateau between 0.25 and 0.45 T 
emerges in the magnetization curve at 1 K. 
The plateau definitely suggests that
an ice-rule obeying spin configuration is more stable than the three-in 
one-out spin configuration at lower temperatures and at lower fields. 
Such plateau is also reported in Ho$_{2}$Ti$_{2}$O$_{7}$ 
at 1.8 K between 0.4 and 1.6 T \cite{Cor1}. 
In Ho$_{2}$Ti$_{2}$O$_{7}$, the appearance of the plateau 
clearly corresponds to the temperature of the maximum in the specific heat. 
This characteristic temperature is 1.97 K 
in Ho$_{2}$Ti$_{2}$O$_{7}$ \cite{Bra1} and corresponds to 
the onset of strong spin-ice correlations in the material. 
The corresponding characteristic temperature in Dy$_{2}$Ti$_{2}$O$_{7}$ 
is 1.2 K \cite{Ram1}. 
The results that a clear plateau appears at 1 K in our calculations 
but not above 1.8 K in our experiments 
as well as our calculations are reasonable 
compared with the characteristic temperature for Dy$_{2}$Ti$_{2}$O$_{7}$. 

In Fig. 5 we show the temperature dependence of reciprocal dc suscpeptibility 
for various magnetic fields along the [100] axis. 
The solid line denotes the Curie-Weiss fitting between 16 and 40 K 
for the data at 0.01 T:
\begin{eqnarray}
\chi ({\it T}) &=&
\frac{{\it N_{\rm A}}p_{\rm eff}^{2}\mu_{\rm B}^{2}/3k_{\rm B}}{{\it
T}-{\it
\theta_{\rm CW}}},
\end{eqnarray}
where $\chi \equiv {\it M/H}$ is magnetic susceptibility,
${\it N_{\rm A}}$ is the Avogadro number,
$p_{\rm eff}$ is effective Bohr magneton, and
$k_{\rm B}$ is the Boltzmann constant.
Again, ${\it H}$ has been corrected for the demagnetization factor.

\begin{figure}
\begin{center}
	\epsfxsize=8cm
	\epsfbox{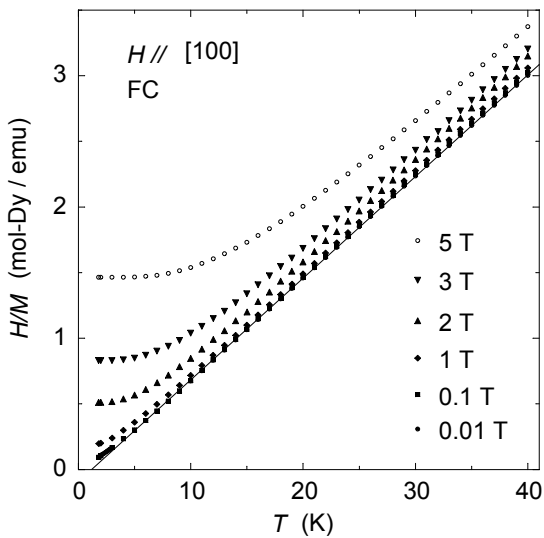}
\end{center}
\caption{
Temperature dependence of reciprocal dc suscpeptibility
for various magnitudes of magnetic field along the [100] axis.
}
\end{figure}

For a given temperature, the ratio ${\it H/M}$ increases 
with an increasing magnetic field. 
This is because the magnetization at higher fields saturates 
compared with that at lower field especially at lower temperatures. 

We obtained positive Curie-Weiss temperatures, 1.16(5), 1.31(5) and 1.28(5) K 
from the data at 0.01 T along the [100], [110] and [111] axes, respectively. 
This is consistent with the existence of 
the effective nearest-neighbor FM interaction, mainly originating from 
large dipolar interaction between Dy ions, and consistent 
with the previous results using polycrystalline samples \cite{Ram1,Bra2}. 
Moreover, this provides further validity in using the value 
for the effective FM interaction used in our calculations, 
${\it J}_{\rm eff} \equiv J_{\rm nn}+D_{\rm nn}= 1.11$ K \cite{Not3}. 

We also obtained the effective Bohr magneton 10.2(1), 10.0(1) and 10.0(1) 
$\mu_{\rm B}$/Dy for the [100], [110] and [111] axes, respectively, 
from the Curie-Weiss fitting. 
These values are nearly equall to the value 
$p_{\rm eff} \equiv g_{\it J}\sqrt{{\it J}({\it J}+1)} = 10.6$ 
$\mu_{\rm B}$/Dy expected for the completely localized spin of Dy$^{3+}$ ion.

\subsection{Ac Susceptibility and Search for the 180 mK Transition}

Considering the above excellent agreement between our experiments 
and calculations, we may expect the realization of the theoretically 
predicted first-order phase transition in the real material \cite{Mel1}. 
In Fig. 6, we show the ac susceptibility $\chi_{\rm ac}$ down to 60 mK. 
In order to keep the sample in zero dc field, 
we surrounded the sample coils with high-permeability paramagnetic tube. 
An ac field of 0.05 mT was applied along the [100] axis. 
This amplitude satisfies the condition, 
${\it H}_{\rm ext}/{\it D}_{\rm nn} \sim 2 \times 10^{-4} \ll 0.08 \sim 
{\it T}_{\rm c}/{\it D}_{\rm nn}$, suggested in Ref. \ref{Rog1}. 
We measured $\chi_{\rm ac}$ at 1 kHz on cooling and 
at 17 Hz on warming after keeping the crystal at 60 mK for 1 hour. 
We normalized the data by dividing the raw data by the measuring frequecy. 

\begin{figure}
\begin{center}
	\epsfxsize=8cm
  	\epsfbox{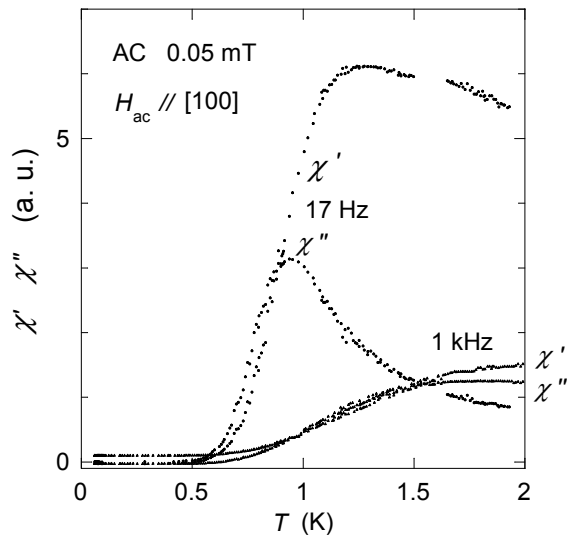}
\end{center}
\caption{
Ac susceptibility measured down to 60 mK.
}
\end{figure}

The in-phase component $\chi'$ measured at 17 Hz exhibits a broad peak 
at around 1.2 K and steep decrease below this temperature. 
In addition, it diminishes below about 0.5 K. 
$\chi'$ at 1 kHz and the out-of-phase, dissipasive components $\chi''$ 
at both frequencies also exibit the similar temerature dependence. 
These data are consistent with those 
obtained from polycrystalline Dy$_{2}$Ti$_{2}$O$_{7}$ \cite{Mat2}. 

The absence of any change in the susceptibilities below 0.5 K 
indicates that the spins on each tetrahedron in the system become 
firmly locked in a random ice-rule configuration at about 0.5 K, 
below which the specific heat also diminishes as shown in the previous 
study by Ramirez $et$ $al$. \cite{Ram1}. 
Further, our data suggest that 
the predicted first-order phase transition does not occur, 
at least within the limited experimental measurement time scale. 
Because the `freeze in' temperature is well 
above the predicted ordering temperature of 0.18 K, it is plausible that 
there is insufficient local spin dynamics left in the system that 
can promote the ordering associated with the nonlocal dynamics 
of spin loop move in the Monte Carlo simulations of Ref. \ref{Rog1}. 
We therefore speculate that the system has essentially completely 
fallen out of equilibrium at around 0.5 K; 
below this temperature one does not expect to recover equilibrium 
in a time that is reasonably accessible by experiment. 

The temperatures of the maxima in $\chi'$ and $\chi''$ increase 
with increasing frequencies. 
The dependence is ascribable to long magnetic relaxation-time 
of spin lock-in formation as Matsuhira {\it et al.} discussed 
in their previous polycrystalline study \cite{Mat2}. 
This also supports our interpretation of 
the absence of ordering down to 60 mK. 
In further studies, it may be possible that slight dilution of Dy 
with non-magnetic elements such as Y may bring the static ordering, 
in analogy to the transition in water ice from the I$_{\rm h}$ phase 
to the long-range ordered crystal I$_{\rm XI}$ phase 
via substitution with KOH \cite{Yam1}.

\section{Conclusion}

In conclusion, we reported magnetization and ac susceptibility 
of Dy$_{2}$Ti$_{2}$O$_{7}$ using single crystals. 
The clear magnetic anisotropy indicates the strong Ising anisotropy 
and the effective FM interaction between the nearest-neighbor Dy ions, 
as well as the dictating ice-rule at low temperatures. 
These effects have been quantitatively confirmed by both experimental 
and theoretical approaches in the present study. 

No indication of the predicted long-range magnetic ordering was observed. 
This may be due to differences in time scale 
between the dynamics of the real system and the simulation. 
Futher investigations, such as specific heat measurements, 
or dilution with non-magnetic impurities may clarify 
the attainability of the real ground state of Dy$_{2}$Ti$_{2}$O$_{7}$.

%%%%%%%%%%%%%%%%%%%%%%%%%%%%%%%%%%%%%%%%%%%%%%%

\vspace{5mm}
\begin{center}

{\bf ACKNOWLEDGEMENT}

\end{center}

\vspace{5mm}

We acknowledge useful discussion and technical support 
from D. Yanagishima and H. Yaguchi. 
We would also like to thank T. Ishiguro for his support in many aspects. 
We appreciate fruitful discussion 
with K. Matsuhira, S. Rosenkranz and Y. Yasui. 
Finally, we acknowledge B. den Hertog for his contribution to this work. 
One of the authors (H. F.) is supported by JSPS Research Fellowships 
for Young Scientist. 
M.G. acknowledges financial support from 
NSERC of Canada, Research Corporation and the Province of Ontario.

%%%%%%%%%%%%%%%%%%%%%%%%%%%%%%%%%%%%%%%%%%%%%%%%

%\vspace{5mm}
%\begin{center}
%
%{\bf APPENDIX}
%
%\end{center}
%
%\vspace{5mm}

%%%%%%%%%%%%%%%%%%%%%%%%%%%%%%%%%%%%%%%%%%%%%%%%

%%%%%%%%%%%%%%%%%%%%%%%%%%%%%%%%%%%%%%%%%%%


\begin{thebibliography}{99}
\bibitem{Qua1} Quantum Design, model MPMS$_{{\rm 5S}}$: we used a
single-axis rotator to adjust the field direction.
\bibitem{Not1} Although the difference between the experiment and the 
calculations at low fields along the [111] direction is small, 
the difference is evident generally for the lower field region. 
This is most likely due to the difficulty 
in precisely and acurately determining the geometric details 
of the sample for the demagnetization factor. 
The difference becomes less than 3 \% at 2 T, 
since it is unimportant to consider the uncertainty of 
the demagnetization factor at high magnetic fields.
\bibitem{Not2} It was quite difficult to exactly align the field direction 
along the [111] axis with the crystal used. 
Because all the crystal surfaces used in the magnetization measurements 
are not parallel to the [111] axis but parallel to the [100] or [110] axes, 
we could not confirm whether the magnetic field is 
parallel to the [111] axis by eye.
\end{thebibliography}
\end{document}